\documentstyle[epsf,prl,aps,multicol]{revtex}
\begin{document}
\draft
\input{psfig}


\title {High-Frequency Hopping
conductivity of Disordered 2D-system in
the IQHE Regime}

\author{I.L.Drichko, A.M.Diakonov,
V.D.Kagan, I.Yu.Smirnov}
\address{A.F.Ioffe Physicotechnical Institute of RAS
Polytechnicheskaya 26, 194021, St.-Petersburg, Russia}

\author{A.I.Toropov}
\address{Semiconductors Physics Institute of SD RAS
Ak.Lavrentieva 13, 630090,  Novosibirsk, Russia}

\date{\today}

\maketitle

\begin{abstract}
High frequency (hf) conductivity in the form
$\sigma^{hf}=\sigma_1^{hf}-i\sigma_2^{hf}$ was obtained
from the measurement of Surface Acoustic Waves (SAW) attenuation and
velocity (f=30 MHz) in GaAs/AlGaAs heterostructures
($n=1.3-7\cdot 10^{11}cm^{-2}$).
It has been shown that in the Integer Quantum Hall Effect (IQHE) regime
for all the samples at magnetic fields corresponding to the middle of the
Hall plateaus and T=1.5 K, $\sigma_1 / \sigma_2 =0.14 \pm 0.03$.
The ratio $\sigma_1 / \sigma_2=0.15$
points the case when the high-frequency hopping conductivity mechanism
(electronic transition between the localized states formed by "tight" pairs)
is valid \cite{efros85}. Dependencies of $\sigma_1$ and $\sigma_2$ on temperature
and magnetic field is analyzed
width of the Landau band broadened by the impurity random potential is
determined.
\end{abstract}

\bigskip


\begin{multicols}{2}
\section*{Introduction}

If one places a semiconducting heterostructure over a piezoelectric which
SAW is being propagated, the SAW undergoes attenuation associated with the
interaction of the electrons of heterostructure with the electric field of
SAW. This is the basis of the
acoustic method pioneered by Wixforth \cite{Wix1} for the investigation of
GaAs/AlGaAs heterostructures. In the paper \cite{ild971} 
it has been found that
in a GaAs/AlGaAs heterostructure in the IQHE regime the acoustically
measured conductivity $\sigma^{hf}$
does not coincide with the $\sigma^{dc}$ obtained from the direct-current
measurements: $\sigma^{dc}$=0, whereas $\sigma^{hf}$ has a finite value.
This difference was explained by means of a conventional model of electrons
being localized in the IQHE
regime,
therefore the conductivity mechanism for a direct current differs from that
for an alternative current. For the localized electrons in the hopping
conduction regime hf-conductivity can be expressed as a complex value:
$\sigma^{hf}=\sigma_1^{hf}-i\sigma_2^{hf}$ \cite{aleiner}. 
Absorption coefficient
of SAW, $\Gamma$, and the change of SAW
velocity- $\Delta V/V$-can be presented in the way:

\begin{eqnarray}
\Gamma=8.68 \frac{K^2}{2}
kA\frac{(\frac{4\pi \sigma_{1}}{\varepsilon_s V})t(k)}
{[1+(\frac{4\pi \sigma_{2}}{\varepsilon_s V})t(k)]^2+
[(\frac{4\pi \sigma_{1}}{\varepsilon_s V})t(k)]^2},
\label{gam&vel}
\end{eqnarray}

$$
A=8b(k)(\varepsilon_1+\varepsilon_0)\varepsilon_0^2\varepsilon_sexp(-2k(a+d)),
$$

$$
\frac{\Delta V}{V}= \frac{K^2}{2}
A\frac{(\frac{4\pi \sigma_{2}}{\varepsilon_s V})t(k)+1}
{[1+(\frac{4\pi \sigma_{2}}{\varepsilon_s V})t(k)]^2+
[(\frac{4\pi \sigma_{1}}{\varepsilon_s
V})t(k)]^2},
$$

where $K^2$  is the electromechanical coupling coefficient of piezoelectric
substrate, $k$ and $V$ are the SAW wavevector and the velocity
respectively, $a$ is the vacuum gap width of 2DEG, $d$ is the depth of
the 2D-layer, $\varepsilon_1$, $\varepsilon_0$ and $\varepsilon_s$ are the
dielectric  constants of lithium niobate, vacuum and semiconductor
respectively, $b$ and $t$ are complex function of $a$, $d$,
$\varepsilon_1$, $\varepsilon_0$  and $\varepsilon_s$.

The aim of the work is to determine Re$\sigma^{hf}$(H,T) and
Im$\sigma^{hf}$(H,T) in IQHE regime from the $\Gamma$ and $\Delta V/V$ of
SAW measurements ($f$=30MHz, T=1.5-4.2K, H up to 7T) and to analyze
the 2D-electrons localization mechanism.

\section*{The Experimental results and discussion}

Fig.1 illustrates the experimental dependencies of $\Gamma$ and $\Delta V/V$
on H at T=1.5K for the sample 1 ($n=2.7 \cdot 10^{11} cm^{-2}$). As long as
$\Gamma$ and $\Delta V/V$ are determined by the 2DEG conductivity (Eq.(1)),
quantizing of the electro
nspectrum in the magnetic field, leading to the SdH oscillations,
results in similar peculiarities in $\Gamma$ and $\Delta V/V$ of fig.1.
Fig.2a presents the $\sigma_1(T)$ dependencies determined from $\Gamma$ and
$\Delta V/V$ (Fig.1) using Eq.(1)
for H=5.5; 2.7 and 1.8T
($\nu$=2, 4, 6 respectively),
$\nu$=nch/eH is the filling factor. Fig.2b shows the $\sigma_2(T)$
dependencies derived from $\Gamma$ and $\Delta V/V$
for H=5.5, 2.7 and 1.8 T. Fig.3 illustrates the dependencies $\sigma_1$ and
$\sigma_2$ on magnetic field near
H=5.5T ($\nu$=2) at different temperatures. In a number of papers
(see f.e.\cite{furlan}) devoted to the study of magnetoresistance 
in IQHE regime
it was established that in IQHE plateau regions at low T the dominant
conductivity mechanism is
variable range hopping.

HF-conductivity  in this case is determined by the two-site model and is
associated with the electronic transition between localized states of
"tight" pairs of impurities $\sigma_1$. In this case the relation is valid:
\begin{eqnarray}
\frac{Re\sigma}{Im\sigma}=\frac{\sigma_1}{\sigma_2}=
\frac{\pi}{2}
ln\frac{\omega}{\omega_{ph}},
\label{reim}
\end{eqnarray}
where $\omega=2\pi f$ is the frequency of SAW, $\omega_{ph}$ is the
characteristic phonon frequency of the order of $10^{12}-10^{13} sec^{-1}$.
The calculation using Eq.(\ref{reim}) gives $\sigma_1/\sigma_2=0.15$ (f=30 MHz).
For all samples it was experimentally found $\sigma_1/\sigma_2=0.14 \pm 0.03$
at T=1.5K and H in the Hall plateaux middle. This fact allows one to suppose
that the mechanism that
determines the $\Gamma$ and $\Delta V/V$  of SAW interacting with localized
electrons is that of the hf hopping conductivity \cite{efros85}.

For the analysis of the dependencies $\sigma_1(H,T)$ and $\sigma_2(H,T) $
we supposed it to be determined by two mechanisms: hf-hopping on the
localized states of impurities and thermal activation to the upper Landau
band. The value of $\sigma_2 $
is proportional to the number of "tight" pairs, which is changed with a
thermal activation processes ($\sigma_2=0$ for delocalized electrons
\cite{ild971}). 
So the dependence $ \sigma_2 (H, T)$ reflects a change of number
of these pairs.

The dependence
$\sigma_1(H)$ is similar to behaviour of $\sigma^{dc}$ when it makes
SdH-oscillations, but $\sigma_1 > \sigma^{dc}=0$. $\sigma_1(T)$ can be
presented as a sum of two terms: $\sigma_1=\sigma_1^h+\sigma_1^a$, where
$\sigma_1^h=0.11 \cdot \sigma_2^h$ is
hf-conductivity,
$ \sigma_1^a=\sigma_0 \cdot exp(-\Delta E/kT)$
is the conductivity on upper Landau band due to the electrons activated from
the states on Fermi level where the activation energy
$\Delta E=\hbar \omega_c/2-C/2$, ($\hbar \omega_c$ is the cyclotron
frequency, $C$
is the Landau band width).
From the plotted dependence $ln(\sigma_1-\sigma_1^h)$ on 1/T
the $\Delta E$ was found for H=5.5, 2.7, 1.8T. One can see (inset of Fig.2a)
that $\Delta E$ is linear function of H with the slope $0.5\hbar \omega_c$
what allow to derive the width of Landau band, broadened,
probably, by
the impurity random potential; the width is appeared to be $C$=2meV.

\section*{Acknowledgements}

The work was supported by RFFI (N 98-02-18280) and Minnauki (N 97-1043).

\end{multicols}

\newpage

\begin{figure}
\centerline{\psfig{figure=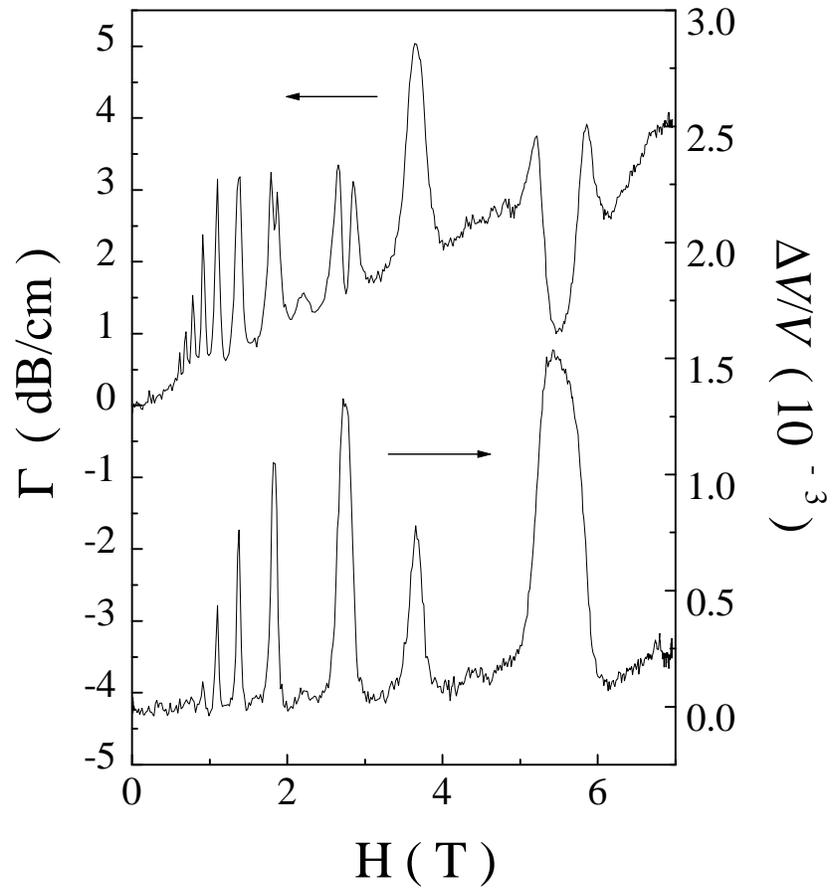,width=15cm}}
\caption{
The experimental dependence of $\Gamma$ and $\Delta V/V$
on H at T=1.5K, ( f=30MHz).}
\end{figure}

\newpage

\begin{figure}
\centerline{\psfig{figure=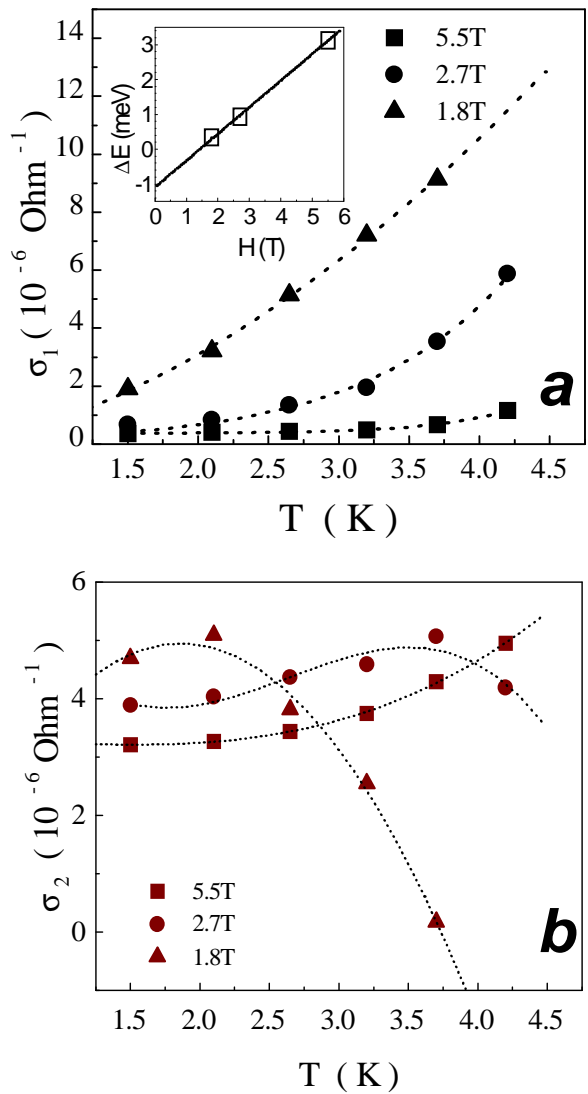,width=15cm}}
\caption{
(a) The dependence of $\sigma_1(T)$ near $\nu$=2; 4;6.
Inset: Activation energy plotted vs. H.
(b) The $\sigma_2(T)$ dependence near the filling factors
$\nu$=2; 4; 6, (f=30MHz).
}
\end{figure}

\newpage

\begin{figure}
\centerline{\psfig{figure=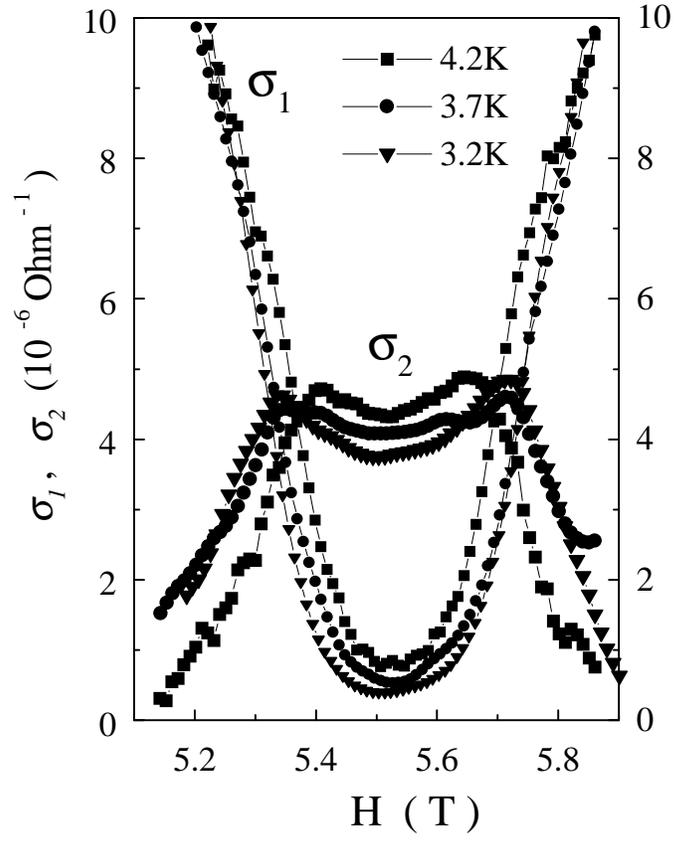,width=15cm}}
\caption{
The $\sigma_1(H)$ and $\sigma_2(H)$ at different temperatures
near the filling factor $\nu$=2.
}
\end{figure}

\end{document}